\documentclass[twocolumn,showpacs,preprintnumbers,amsmath,amssymb]{revtex4}

\usepackage{graphicx}
\usepackage{dcolumn}
\usepackage{bm}

\begin{document}

\title{Exchange induced charge inhomogeneities in rippled neutral  graphene}

\author{L. Brey$^{1}$ and
 J. J. Palacios$^{1,2}$. }
\affiliation{ (1)Instituto de Ciencia de Materiales de Madrid
(CSIC), Cantoblanco 28049, Spain
\\ (2)Departamento de F\'{\i}sica Aplicada, Universidad de Alicante, San Vicente del Raspeig, E-03690 Alicante
Spain}

\date{\today}

\begin{abstract}
A new mechanism that  induces charge density variations in
corrugated graphene is proposed. Here it is shown how the interplay
between lattice deformations and exchange interactions can induce
charge separation, i.e., puddles of electrons and holes, for
realistic deformation values of the graphene sheet. The induced
charge density lies in the range of $10^{11}-10^{12}$ cm$^{-2}$,
which is compatible with recent measurements.
\end{abstract}
\pacs{73.21.-b,73.20.Hb,73.22-f} \maketitle

\emph{Introduction.} The ability to isolate and manipulate graphene
-single layers of carbon atoms packed in a honeycomb lattice- has
opened and boosted the experimental study of the properties of two
dimensional (2D) massless Dirac
fermions\cite{Novoselov_2004,Novoselov_2005,Zhang_2005}. The
existence of a strictly 2D crystal structure is puzzling in itself,
as, according to theory and experiments, perfect 2D crystals can not
exist in the free
state\cite{Mermin_1968,Landau_book,Novoselov_2005a}.

Recently, transmission electron microscopy
experiments\cite{Meyer_2007} have resolved that suspended graphene
sheets are not perfectly flat but exhibit microscopic roughness or
ripples  such that the surface normal varies by several degrees and
the out of plane deformation can reach 1nm. This deformation
corresponds to a rather large strain, 1$\%$ , but is sustainable
without plastic deformations or generation of
defects\cite{Meyer_2007}. Also, layers placed on SiO$_2$ seems to
follow the corrugation of the
substrate\cite{Stolyarova_2007,Ishigami_2007}. The height and width
of these ripples are consistent with models which allow the carbon
ions to form different types of bonds\cite{Fasolino_2007}.


Recent single electron transistor based
experiments\cite{Martin_2007} have evidenced the existence of
electron and hole puddles of densities $\sim 10^{10} -10^{11}$
cm$^{-2}$ in the vicinity of the neutrality point. The existence of
these puddles could be simply related to the presence of a disorder
potential induced by the
substrate\cite{hwang-2007-98,Shklovskii_2007}.
An alternative to this explanation bears on the existence of ripples
which leads to a modulation of the hopping amplitudes between carbon
atoms. This modulation affects the electronic structure in two-fold
manner. Firstly, the modulation induces an effective magnetic field
which changes locally the density of states, but does not separate
charge\cite{Castro-Neto_2007}. In this respect
it has been argued in Ref. \onlinecite{Guinea_2007} that a
one-dimensional deformation of the graphene sheet will form zero
energy  Landau levels corresponding to an effective magnetic field
of tens of  Teslas. This will increase the compressibility of the
system and eventually would induce electronic phase separation.
Secondly, second-neighbor hopping changes can induce a potential on
the carriers which does separate charge\cite{Castro-Neto_2007}.
An additional source of change in the local density of states has
been suggested to come from local Fermi velocity changes induced by
the curvature associated to the ripples\cite{Juan_2007}. This effect
induces charge inhomogeneities in doped graphene, but, in the
presence of electron-hole symmetry, can not explain the existence of
electron-hole puddles in undoped graphene.

In this paper we propose that because of the dependence of the
exchange energy on the density of carbon  atoms, the strain
modulation produced by the ripples  induces a charge modulation in
undoped graphene. In order to emphasize and highlight  the effect of
the exchange on the charge inhomogeneity we have disregarded in this
work the dependence of the hopping parameter on next neighbor
distance. This is a justified assumption as in graphene a variation
of the hopping amplitude can be described by means of a fictitious
vector potential which, in general, does not induce significant
changes in the density of states\cite{Castro-Neto_2007}.  In other
words, the degeneracy of the Landau levels and their contribution to
the compressibility\cite{Guinea_2007} will be strongly suppressed if
a randomly curved graphene sheet\cite{Meyer_2007,Fasolino_2007} is
considered.

\begin{figure}
  \includegraphics[clip,width=8cm]{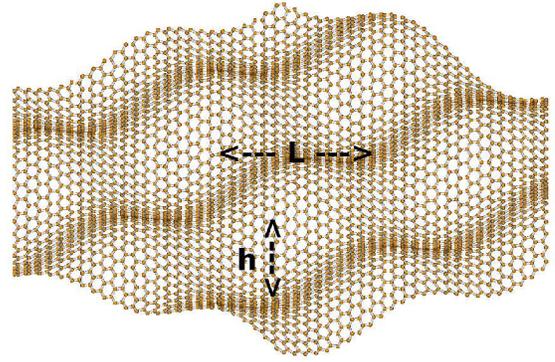}
  \caption{($Color$ $online$) Schematic view of graphene with periodic ripples.
  $L$ denotes the lateral extent of the ripples and $h$ the out of plane displacements of the Carbon atoms.}
   \label{ripples}
\end{figure}

In the following we first point out how the existence of ripples in
the graphene surface  creates a modulation of the distance between
first neighbor carbon atoms. Then, within the Hartree-Fock
approximation, we obtain the total energy of the system as a
function of the distance between first neighbors atoms, $d$, and of
the extra charge per carbon atom, $\delta$. Finally, we show that a
long-wavelength modulation of the lattice parameter in the graphene
sheet can induce a charge density modulation in undoped graphene.

\emph{Distance between carbon atoms}. To simulate the experimental
electron diffraction patterns\cite{Meyer_2007}, it is needed to
assume a ratio $L/h \approx 10$ between the lateral size of the
ripples, $L$, and the out of plane displacement, $h$ (see Fig.
\ref{ripples}). A prudent estimate for the typical lateral size of
the crumpling  has been estimated to be  between 2 and 20
nm\cite{Meyer_2007}. In our calculation we consider the following
out of phase modulation,
\begin{equation}
z(\textbf{R}_i) = h \left [ \sin {(G \, R_{x,i})}+ \sin {(G \,
R_{y,i})} \right ]
\end{equation}
where $G=2 \pi / L $ and the vectors $\textbf{R}_i $ are the
position of the carbon atoms in a perfectly flat graphene sheet. The
average distance between an atom and its three first neighbors
depends on the position of the considered atom and is given by
\begin{equation}
d (\textbf{R} _i) \simeq \tilde{d} \,  \left ( 1 + \frac {h ^2 G
^2}{8 } [ \cos  { (2 \, G \, R_{x,i})}  + \cos  {( 2\, G \,
R_{y,i})}  ] \right ), \label{dist_mod}
\end{equation}
being $\tilde{d}= d_0 (1+h^2 G^2 /4)$ the average distance between
carbon atoms in the presence of the ripple  and $d_0$ the
equilibrium distance between the C atoms in flat graphene.
Expression (\ref{dist_mod}) indicates that a modulation in the out
of plane position  of the carbon atoms implies a modulation in the
average distance between them.

\emph{Electronic structure and kinetic energy.} The electronic
active states of graphene are the bands formed by the carbon $p_{\rm
z}$ orbitals which are ordered in a honeycomb lattice (triangular
lattice with two atoms per unit cell). The band structure is
accurately described by means of  a first neighbors tight-binding
Hamiltonian with a unique hopping parameter $t$. Because the
bipartite character of the honeycomb lattice, the hamiltonian  in
reciprocal space is represented by a 2x2 matrix. In undoped graphene
there is  one electron per carbon atom  and the conduction and
valence bands touch at two non-equivalent points of the Brillouin
zone that are called Dirac points.

For wavevectors near the Dirac  points the electronic structure can
be described by a Dirac Hamiltonian of the form
\begin{equation}
H_{kin} = \hbar v_F  \, \textbf{k} \cdot \bf{ \sigma}
\label{H_Dirac}
\end{equation}
where  $\bf{\sigma}$ are the Pauli matrices, $\bf{k}$ is the
electron momentum measured with respect the Dirac point and the
Fermi velocity only depends on the hopping energy $t$ and the next
neighbors distance   $d_0$,
\begin{equation}
\hbar v _F =\frac {3} 2 \, t \, d_0 \, \, \, .
\end{equation}
The eigenvalues and eigenvectors of Hamiltonian Eq.\ref{H_Dirac} are
\begin{equation}
\varepsilon _{\pm}( {\bf k} ) =  \pm \hbar v_F |{\bf k}| \, \, \, \,
\textrm{and} \, \, \,  \psi _{\pm} (\textbf{k} )=\frac {e ^{i
\textbf{k} \textbf{r} }}{\sqrt{2}} \left (
\begin{array} {c} 1
\\ \pm e ^{i \theta _{\bf k}} \end{array} \right ) \, ,
\end{equation}
where $\theta _{\bf k} = \arctan k_x/k_y$.

The use of the continuum Dirac Hamiltonian for describing the
properties of graphene requires the introduction of a maximum value
of  momenta, $k_c$, which is chosen to keep the number of states in
the Brillouin zone fixed, i.e., $g_d \pi k_c ^2 = (2 \pi ) ^2/S_0$.
Here $g_d=2$ is the Dirac points degeneracy and $ S_0 =3 \sqrt{3}
d_0  ^2 /2$ is the unit cell area.

When doping graphene with  electrons or holes the extra carriers
form a Fermi sea with Fermi wavevector $k_F= \sqrt{ 4 \pi |n| /(g_d
g_s )}$, where $n$ is the 2D density of added charge and $g_s=2$ is
the spin degeneracy. The kinetic energy of the system per carbon
atom is given by
\begin{eqnarray}
E_{kin} (\delta) & = & S_0 \frac {\hbar} { 3 \pi}  v _F \left ( -
k_c ^3+k_F ^3
\right )\nonumber \\
& = & -t \left ( \frac {  \pi }{6 \sqrt{3}} \right ) ^{1/2} \left [2
^{3/2} - |\delta| ^{3/2} \right ] \label{e_kin}
\end{eqnarray}
where $\delta$ is the extra charge per carbon atom with respect the
intrinsic situation. Note that due to the linear dispersion of the
bands, when we discard the variation of $t$ on distance, the kinetic
energy per carbon atom does not depend on the distance between
atoms. From the expression of the kinetic energy, Eq.(\ref{e_kin}),
there is  always a kinetic energy  cost associated to modulating the
charge in undoped graphene.

\emph{Exchange energy}. The exchange contribution to the total
energy per carbon atom has the
form\cite{Peres_2005a,Hwang_2007,Barlas_2007}
\begin{equation}
E_{EX} = - \frac {g_s g_d } 4 S_0  \sum _{s,s',{\bf k},{\bf q}} v
(q) F_{s,s'} ({\bf k},{\bf k} + {\bf q}) n_s({\bf k}) n_{s'}({\bf
k}+ {\bf q}), \label{EX_Energy}
\end{equation}
where $s$ and $s'$ is the band index ($\pm 1$),  $v(q)= 2 \pi e ^2 /
\epsilon q$ the 2D Fourier transform of the Coulomb interaction,
$\epsilon$ is the dielectric constant of the system, $n_s ({\bf k})$
is the Fermi occupation function of the state $\{s,{\bf k}\}$ and
$F_{s,s'}({\bf k},{\bf k}+{\bf q})$ is the square of the overlap
between the wavefunctions, $\psi _{s} (\textbf{k} )$ and $\psi _{s'}
(\textbf{k}+ \textbf{q} )$\cite{wunsch_2006,brey_2007}
\begin{equation}
F_{s,s'}({\bf k},{\bf k}+{\bf q})= \frac{1}{2} (1 + s s' \cos
{\theta}), \label{overlap}
\end{equation}
with  $\theta$ the angle between the wavevectors ${\bf k}$ and ${\bf
k} + {\bf q}$. For Coulomb interaction, the factor (\ref{overlap}),
makes the exchange interaction to be larger between states in the
same band. In  the expression Eq. (\ref{EX_Energy}) we have
neglected the exchange energy between electrons belonging to
different Dirac cones. This is appropriated in the long wavelength
approximation because the difference in momentum between states
coming from different Dirac points is very large.

Following the notation of Ref. \onlinecite{Peres_2005a} the exchange
energy per carbon atom can be written as
\begin{eqnarray}
E_{EX}(d, \delta) = -\frac 1 {16\pi ^2} \frac {e^2}{ \epsilon d}
\left( \frac {2 \pi }{\sqrt{3}} \right )^{3/2} \times \nonumber \\
\left [ 2^{3/2} R_1(1)  + |\delta|^{3/2} \, R_1 (1) +  \delta \,
2^{3/2} \, R_2 \left (\sqrt{ \frac{|\delta|}{2}} \right ) \right ]
\label{e_ex}
\end{eqnarray}
where the functions $R_n (a)$ are defined in Ref.
\onlinecite{Peres_2005a}. In the limit of small extra charge one can
approximate
\begin{eqnarray}
R_1(1) & \simeq & 3.776 \nonumber \\
R_2 \left (\sqrt{ \frac{|\delta|}{2}} \right ) & \simeq & \frac
{\pi}{3} \left [3+\sqrt{ \frac{|\delta|}{2}}  \ln {\sqrt{
\frac{|\delta|}{2}}} + ...\right ] \, \, .
\end{eqnarray}
In Eq. \ref{e_ex} the first term corresponds to the exchange energy
of the full occupied valence band, the second term the exchange
energy of the extra electrons or holes, and the last term is the
variation of the exchange energy because the interaction between the
extra carriers and the valence band electrons. Note that the last
term in Eq. \ref{e_ex} changes sign with the electron/hole character
of the extra carriers.

\emph{Coupling between lattice deformation  and charge.} Expressions
Eq. (\ref{e_kin}) and Eq. (\ref{e_ex}) have been obtained for a
uniform system. For long wavelength modulations of $d$ and $\delta$,
the total energy per carbon atom  can be written as
\begin{equation}
E_{T} = \frac 1 S \int d \textbf{r} \left [ E_{kin} (\delta({\bf
r})) + E_{EX}(d({\bf r}), \delta({\bf r})) \right ] \, \, .
\label{Total_Energy}
\end{equation}
This equation couples a modulation of the first neighbors distance
between carbon atoms with a modulation of the electron density.

In undoped graphene a modulation  $d(\textbf{R})$ of the form
described in Eq.\ref{dist_mod} will produce, in lineal regime, a
modulation of the electric charge of the form,
\begin{equation}
\delta  (\textbf{R} ) =\delta _1 \, [ \cos  {( 2 \, G \, R_{x,i})} +
\cos  { (2\, G \, R_{y,i})}  ] \label{charge_mod}
\end{equation}
where the amplitude of the charge modulation, $\delta _1$, is
obtained by minimizing the total energy  Eq.\ref{Total_Energy}. To
lowest order, the charge amplitude $\delta_1$ only depends on the
ratio $h/L$ and on a dimensionless coupling constant defined as
\begin{equation}
g =\frac{e^2 /\epsilon _0}{\hbar v_F} \, \, . \label{g}
\end{equation}
The constant $g$  indicates the ratio between the Coulomb and the
kinetic energy of the electron system.

\begin{figure}
  \includegraphics[clip,width=8cm]{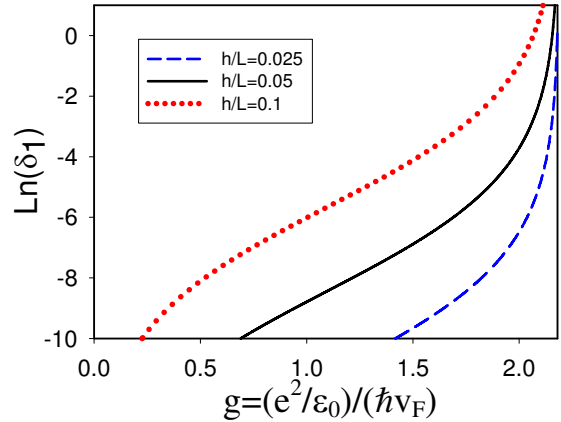}
  \caption{($Color$ $online$) Amplitude of the charge modulation per Carbon
  atom $\delta _1$
  as function of  the
  coupling constant, for different values of the ratio $h/L$.}
   \label{Figure2}
\end{figure}

For values larger  than a critical coupling constant $g_c \sim
2.18$, the total energy does not present a minimum as function of
$\delta _1$. This indicates that the system is unstable against more
correlated phases and therefore  the lower order calculation fails.
This value of the coupling constant is considerably smaller than the
critical value of $g$ ($g \sim 5.3$) needed for the occurrence of
instabilities towards highly correlated states in flat
graphene\cite{Peres_2005a}. For  values of $g < 2.18$ the total
energy  presents a minimum and the electronic system reacts to the
lattice deformation by modulating the charge according to Eq.
\ref{charge_mod}. In Fig.\ref{Figure2} we plot the value of
$\delta_1$ that minimize the total energy as function of the
coupling constant for different values of the ratio $h/L$.

The  distance between carbon atoms in graphene and the hopping
amplitude are quantities well established; $t\cong2.5eV$ and
$d_0$=1.42$\AA$. Therefore a realistic estimation of the coupling
constant, $g$, only requires the knowledge of the value of the
dielectric constant.  This should take account both the effect of
the screening current in the substrate as well as the weak intrinsic
screening in graphene\cite{Gonzalez_1999}. We estimate that
depending whether the graphene sheet is freely suspended or it is
placed on top of an insulator, the dielectric constant would take
values in  the range $3< \epsilon <4$. For these values of
$\epsilon$, the coupling constant takes values in the range
$0.66<g<0.88$.
For these values of $g$, we can see in Fig.\ref{Figure2} that the
use of the lower order coupling between $h$ and $\delta _1$ is
appropriated.

For an intermediate coupling constant, $g=0.75$, we obtain $\delta_1
\simeq 9 \left (\frac {h^2} {L^2} \right )^2$ which, for values of
the ratio $h/L$ in the range 0.05 to 0.1, implies values of $\delta
_1$ in the range 5.6$\times 10 ^{-5}$ to 0.9$\times 10 ^{-4}$. This
corresponds to density modulations in the range 2.25$\times 10 ^{11}
$ cm$^{-2}$ to 3.6$\times 10 ^{12}$ cm$^{-2}$. From this estimate we
conclude that a modulation of the out of plane position of the
carbon atoms of amplitude $h \sim 1-2 $ nm in a lateral size $L \sim
10-20$nm induces a modulation in the charge density of order $10
^{11}$ cm$^{-2}$. This magnitude of the charge modulation agrees
with the density of charge in the electron-hole puddles observed by
single electron transistor based experiments\cite{Martin_2007}.
Interestingly, in suspended graphene the contribution of the
substrate to the dielectric constant is practically suppressed. In
this case the coupling constant  becomes larger than in the case of
graphene placed on a dielectric. Therefore  we expect a higher
density modulation in suspended graphene than the observed in
graphene on SiO$_2$.

It is also pertinent to estimate the importance of the Hartree
repulsion on the values of the charge density modulation. The
Hartree energy per carbon atom takes the form
\begin{eqnarray}
E_H &=& \frac {S_0}{8 \pi ^2} \sum _{\textbf{G}'} \frac {2 \pi e
^2}{\epsilon G'} n(\textbf{G}')n(-\textbf{G}') \nonumber \\
&=& \frac {1}{8 \pi ^2 3 \sqrt{3}} \delta _1 ^2 \frac {L} {d_0}
\frac {e^2}{\epsilon d_0} \, \, \, .
\end{eqnarray}
In the above expression $n(\textbf{G}')$ is the $\textbf{G}'$
component Fourier transform of the charge. For the values of
dielectric constant and $L$ we obtain that the Hartree energy is
much smaller than the kinetic and exchange energy and the values
obtained for the charge density modulation are practically
unaffected by the Hartree repulsion.

On the top of the effect discussed in this work one should also
consider the influence of the hopping dependence on lattice strain
discussed in Refs. \onlinecite{Castro-Neto_2007,Guinea_2007}.
Realistic estimates of the influence of second-neighbor hopping
variations indicate that these are typically smaller that the one
discussed here. In any case, both effects are compatible. Regarding
the formation of Landau levels and their influence on the
compressibility, we have verified numerically\cite{Juanjo_2007} that
that density of states close the neutrality point is not seriously
affected by the effective magnetic fields when crumpled graphene
with uncorrelated ripples is taken into consideration. This
contrasts with the strong influence that periodic ripples exert on
the density of states near the Dirac point\cite{Guinea_2007}.

\emph{Summary.} We notice that exchange interaction between carriers
produces a coupling between a modulation of the distance between
first neighbors atoms in graphene and a charge density modulation.
This mechanism connects the presence of ripples in undoped graphene
with the existence of electron hole puddles of density up to
$10^{12} cm^{-2}$. These densities are of the same order than the
observed recently in single electron transistor
measurements\cite{Martin_2007}. In suspended graphene the absence of
substrate makes the exchange energy stronger and we predict that
this increase  will produce a higher modulation of the charge
density.

We acknowledge useful discussions with F. Guinea, M. A. H.
Vozmediano, A. Cortijo and C. Tejedor. This work has been
financially supported by MEC-Spain (Grants FIS2004-02356 and
MAT2006-03741) and by Generalitat Valenciana (Grants ACOMP07/054 and
GV05-152). This work has also been partly funded by FEDER funds.


\end{document}